\documentclass[twoside,12pt]{article}
\usepackage{amsmath}
\usepackage{amssymb}
\usepackage{latexsym}
\usepackage{epsfig}
\usepackage{cite}
\newlength{\dinwidth}
\newlength{\dinmargin}
\newenvironment{proof}{\medskip \noindent 
            {\bf Proof.}}{ \hfill $\square$ \medskip}
\newtheorem{lemma}{Lemma}

\newtheorem{Proposition}{Proposition}
\newtheorem{Def}{Definition}

\def\bbbr{{\rm I\!R}} 

\def\cA{{\cal A}}
\def\cB{{\cal B}}
\def\cC{{\cal C}}
\def\cH{{\cal H}}
\def\cM{{\cal M}}
\def\cN{{\cal N}}
\def\cP{{\cal P}}

\def\PN{{\cal P} ({\cal N} )}
\def\c{{\bot}} 
\def\vagy{\vee}
\def\es{\wedge}
\def\C{$C^{\ast}$--}

\def\idty{{\leavevmode\hbox{\rm 1\kern -.3em I}}}

\begin{document}

\title{Local Primitive Causality \\ and the \\ Common Cause Principle \\
in\\ Quantum Field Theory}  

\author{{Mikl\'os R\'edei} \\
Department of History and Philosophy of Science  \\  
Lor\'and E\"otv\"os University\\ 
P.O. Box 32, H-1518 Budapest 112, 
Hungary\thanks{e-mail: redei@ludens.elte.hu}\\
\vphantom{X}\\
{Stephen J.\ Summers } \\
Department of Mathematics, University of Florida,\\
Gainesville FL 32611, USA\thanks{e-mail: sjs@math.ufl.edu}}

\date{August 3, 2001}

\maketitle

\abstract{If $\{\cA(V)\}$ is a net of local von Neumann algebras satisfying 
standard axioms of algebraic relativistic quantum field theory and 
$V_1$ and $V_2$ are spacelike separated spacetime regions, then 
the system $(\cA(V_1),\cA(V_2),\phi)$ is said to satisfy the Weak 
Reichenbach's Common Cause Principle iff for every pair of 
projections $A\in\cA(V_1)$, $B\in\cA(V_2)$ correlated in the normal 
state $\phi$ there exists a projection $C$ belonging to a von Neumann algebra 
associated with a spacetime region $V$ contained in the union of the backward 
light cones of $V_1$ and  $V_2$ and disjoint from both $V_1$ and $V_2$, a 
projection having the properties of a Reichenbachian common cause of the 
correlation between $A$ and $B$. It is shown that if the net has the local 
primitive causality property then every local system 
$(\cA(V_1),\cA(V_2),\phi)$ with a locally normal and locally faithful state 
$\phi$ and open bounded $V_1$ and $V_2$ satisfies the Weak Reichenbach's 
Common Cause Principle.} 

\section{Introduction}

     An operationally motivated and mathematically powerful approach
to quantum field theory is algebraic quantum field theory (AQFT) (cf.
\cite{Haag1992}). Although in its axiomatic nature it includes models
which have no physical significance at all, it does subsume all
physically interesting models, as well, for the assumptions made
in AQFT are ordinarily of such a nature that they are the desiderata
of any reasonable quantum field theory. The initial data in AQFT are a 
collection $\{\cA(V)\}$ of algebras indexed by a suitable set of open 
subregions of the space--time of interest, $\cA(V)$ understood as
being generated by all
the observables measurable in the spacetime region $V$, and a state
$\phi$ on these algebras, understood as representing the preparation
of the quantum system under investigation. It is remarkable that 
it is not necessary to make a specific choice of observables and state
({\it i.e.} to choose a particular model) in order to establish results 
of physical interest --- many such results follow from quite general 
assumptions.

     A characteristic feature of (relativistic) local algebraic quantum
field theory is that it predicts correlations between
projections $A,B$ lying in von Neumann algebras $\cA(V_1),\cA(V_2)$ 
associated with spacelike separated spacetime regions
$V_1,V_2$. Typically, if $\{\cA(V)\}$ is a net of local algebras in a
vacuum representation, then there exist many normal states $\phi$ on 
$\cA(V_1\cup V_2)$ such that $\phi(A\es B)>\phi(A)\phi(B)$ for
suitable projections $A \in \cA(V_1)$, $B \in \cA(V_2)$ (this will be 
explained below).

     According to a classical tradition in the philosophy of science,
such probabilistic correlations are always signs of causal
relations. More precisely, this position is typically formulated in
the form of what has become known as {\em Reichenbach's Common Cause
Principle}. This principle asserts (cf. \cite{Salmon1984}) that if two
events $A$ and $B$ are correlated, then the correlation between $A$
and $B$ is either due to a direct causal influence connecting $A$ and
$B$, or there is a third event $C$ which is a common cause of the
correlation. The latter means that $C$ satisfies four simple
probabilistic conditions which together imply the correlation in
question. (These conditions will be given below.)

     The self--adjoint elements of the local von Neumann algebras $\cA(V)$
associated with spacetime regions $V$ are interpreted in AQFT as the
mathematical representatives of the physical quantities observable in
region $V$. Since, in particular, projections in the local von Neumann
algebras are interpreted as 0-1--valued observables and the expectation
values of the projections in $\phi$ as probabilities of the events that these
two-valued observables take on the value 1 when the system has been
prepared in state $\phi$, then the above--mentioned correlations
predicted by AQFT lead naturally to the question
of the status of Reichenbach's Common Cause Principle within AQFT.
If the correlated projections belong to algebras associated with
spacelike separated regions, a direct causal influence between them is
excluded by the theory of relativity. Consequently, compliance of AQFT
with Reichenbach's Common Cause Principle would mean that for every
correlation between projections $A$ and $B$ lying in von Neumann
algebras associated with spacelike separated spacetime regions
$V_1,V_2$ there must exist a projection $C$ possessing the probabilistic
properties which qualify it to be a Reichenbachian common cause of the
correlation between $A$ and $B$. However, since observables and hence
also the projections in AQFT must be localized, in the case of the
spacelike correlations predicted by AQFT, one also has to specify the
spacetime region $V$ with which the von Neumann algebra $\cA(V)$
containing the common cause $C$ is associated. Intuitively, the region
$V$ should be disjoint from both $V_1$ and $V_2$ but should not be
causally disjoint from them, in order to leave room for a
causal effect of $C$ on the correlated events. Thus the natural
requirement concerning the region $V$ is that it be contained in the
intersection of the backward light cones of $V_1$ and $V_2$.

     This requirement and the resulting notion of Reichenbachian common
cause in AQFT (Definition \ref{defccqft}) was formulated in a previous
paper \cite{Redei1997} (see also Chapter 8 in \cite{konyvem}),
together with the problem of whether AQFT complies with Reichenbach's
Common Cause Principle, as described. This problem is still open --- 
the present paper does not settle the issue. What we can show
here is less --- we shall prove that if a net $\{\cA(V)\}$ of local
von Neumann algebras satisfies some standard, physically natural
assumptions as well as the so--called {\em local primitive causality}
condition (Definition \ref{primitive}), then every local system
$(\cA(V_1),\cA(V_2),\phi)$ with a locally normal and locally faithful
state $\phi$ and open, bounded spacelike separated 
spacetime regions $V_1,V_2$ satisfies
the Weak Reichenbach's Common Cause Principle, where ``weak'' means
that there exists a region $V$ contained in the {\em union} of the
backward light cones of $V_1$ and $V_2$ such that the local von
Neumann algebra $\cA(V)$ contains a common cause $C$ of the
correlation (Definition \ref{wdefccqft} and Proposition \ref{main}).
We shall show that such states include the states of physical interest
in vacuum representations for relativistic quantum field theories on
Minkowski space. Hence, although the question of whether
general local systems $(\cA(V_1),\cA(V_2),\phi)$ with non-faithful
normal states $\phi$ satisfy the Weak Reichenbach's Common Cause
Principle remains open, we will interpret Proposition \ref{main} as a
clear indication that AQFT is a causally rich enough theory to comply
with the Weak Common Cause Principle -- and possibly also with the
strong one.

\section{Spacelike Correlations in Quantum Field Theory}

In light of the interpretation given to the net $\{\cA(V)\}$ of 
\C algebras $\cA(V)$ (with common identity element) indexed by the
open, bounded subsets $V$ of Minkowski space $M$,\footnote{The results 
presented in this paper can be generalized to suitable nets and states on 
stationary, globally hyperbolic space--times, but we shall content ourselves 
here with illustrating the ideas in Minkowski space.}
the following properties are physically natural requirements
(for further discussion of these axioms, see 
\cite{Haag1992} and \cite{Horuzhy1990}):
\begin{description}
\item[(i)]   Isotony: if $V_1$ is contained in $V_2$, then $\cA(V_1)$ 
        is a subalgebra of $\cA(V_2)$; 
\item[(ii)]  Einstein causality: if $V_1$ is spacelike separated from 
$V_2$, then 
every element of $\cA(V_1)$ commutes with every element of $\cA(V_2)$; 
letting $\cA(V)'$ denote the commutant of $\cA(V)$ in $\cA$, this can be
succinctly expressed by $\cA(V_1) \subset \cA(V_2)'$;
\item[(iii)] Relativistic covariance: there is a representation $\alpha$ 
of the identity-connected component $\cP$ of the Poincar\'e group by 
automorphisms on $\cA$ such that $\alpha_g(\cA(V))=\cA(gV)$  for all 
$V$ and $g\in\cP$. 
\end{description} 
The smallest \C algebra $\cA$ containing all the local algebras
$\cA(V))$ is called the quasilocal algebra. This is the algebra on
which $\phi$ is defined as a state. In this paper we shall be interested in 
irreducible vacuum representations, so we shall also assume: 

\begin{description} 
\item[(iv)] Vacuum representation: for each $V$, $\cA(V)$ is a von
Neumann algebra acting on a separable Hilbert space $\cH$ in which
$\cA = \cB(\cH)$ (the set of all bounded operators on $\cH$) and there 
is a distinguished unit vector $\Omega$, and on which 
there is a strongly continuous unitary representation $U(\cP)$ such that
$U(g)\Omega = \Omega$ , for all $g \in \cP$, and 
$$\alpha_g(A) = U(g)AU(g)^{-1} \quad , \quad {\rm for \; all} \quad
A \in \cA \quad ,$$
as well as the spectrum condition --- the spectrum of the self--adjoint
generators of the strongly continuous unitary representation $U(\bbbr^4)$
of the translation subgroup of $\cP$ (which has the physical
interpretation of the global energy--momentum spectrum of the theory)
must lie in the closed forward light cone.
\end{description}

     The vacuum state $\phi$ on $\cA$ is defined by
$\phi(A) = \langle\Omega,A\Omega\rangle$, for all $A \in \cA$. This
state is thus Poincar\'e invariant: $\phi(\alpha_g(A)) = \phi(A)$,
for all $g \in \cP$ and $A \in \cA$. We shall further assume that
there are sufficiently many observables localized in arbitrarily small 
spacetime regions:

\begin{description} 
\item[(v)] Weak additivity: for any nonempty open region $V$, the set
of operators $\cup_{g \in {\bbbr}^4} \cA(gV)$ is dense in $\cA$ (in the weak
operator topology).
\end{description}

     An immediate consequence of assumptions (i)--(v) is the
Reeh--Schlieder Theorem: for any nonempty open region $V$, the set of
vectors $\cA(V)\Omega = \{ A\Omega \mid A \in \cA(V) \}$ is dense in
$\cH$. In fact, one has a stronger result, which will be useful for
us.  First, we must recall some definitions.  For a spacetime region
$V$, let $V'$ denote the (interior of the) causal complement of $V$,
{\it i.e.} the set of all points $x\in M$ which are spacelike
separated from every point in $V$.  In addition, one says that a
vector $\Phi$ in $\cH$ is analytic for the energy if the function $\mathbb C
\ni z \mapsto e^{zH}\Phi \in \cH$ is analytic, where $H$ is the
generator of the one-parameter group $U(t)\in U(\bbbr^4)$, $t \in \bbbr$,
implementing the time translations. By the spectrum condition, $H$ is
a positive operator. Note that any vector $\Phi$ with finite energy
content --- {\it i.e.} for which there exists a compact set in the
spectrum of $H$ such that the corresponding spectral projection $P$
leaves $\Phi$ fixed, {\it i.e.}  $P\Phi = \Phi$ --- is a vector
analytic for the energy. In particular, the vacuum vector $\Omega$ is
analytic for the energy. And since no preparation of a quantum system
which can be carried out by man can require infinite energy, it is
evident that (convex combinations of) states induced by such analytic
vectors include all of the physically interesting states in this
representation.

\begin{Proposition} {\rm \cite{Borchers}} 
Under the assumptions (i)--(v), for any 
nonempty open region $V$, the set of vectors $\cA(V)\Phi$ is dense 
in $\cH$, for all vectors $\Phi$ which are analytic for the energy.
\end{Proposition}

     Note that assumption (ii) entails that such vectors are also
separating ({\it i.e.} $X \in \cA(V)$ and $X\Phi = 0$ imply $X = 0$)
for all algebras $\cA(V)$ such that $V'$ is nonempty. Hence, 
(convex combinations of) the states $\phi$ induced by analytic vectors are 
faithful ({\it i.e.} $X \in \cA(V)$ and $\phi(XX^*) = 0$ imply $X = 0$)
on each such algebra $\cA(V)$. Such states are said to be locally faithful.
We emphasize: given assumptions (i)--(v), all physically interesting
states will be locally faithful.

     We shall be interested in special bounded regions $V$, called
double cones. Let the point $x \in \cM$ be in the interior of the
forward light cone with apex at $y \in \cM$. Any two such points
determine a double cone as the interior of the intersection of the
backward light cone with apex at $x$ with forward light cone with apex 
at $y \in \cM$. The set of double cones forms a basis for the topology
on $\cM$. Double cones are nonempty, as are their causal complements.
We state a further assumption.

\begin{description} 
\item[(vi)] Type of double cone algebras: for every double cone $V$,
the von Neumann algebra $\cA(V)$ is type $III$. \footnote{cf. \cite{KadRing}
for a description of types of von Neumann algebras}
\end{description}

     This assumption appears to be of a different sort than the
previous ones, but it has been verified in many models by direct
computation, and it has been derived from general principles (for
example, the existence of a scaling limit \cite{Fredenhagen}).  So,
typically, double cone algebras in vacuum representations are, in
fact, type $III$, and condition (vi) is not anticipated to be
physically restrictive.

     We formulate our final assumption. For a convex spacetime region
$V$, let $V''=(V')'$ denote the {\em causal completion} (also called
causal closure and causal hull in the literature) of $V$.  One notes
that every light ray running through any given point in $V''$ must
intersect $V$. One should also note that $V = V''$ for every double
cone $V$. There are many ways of formulating a causality condition in
AQFT short of specifying a particular time propagation
(cf. \cite{Haag-Schroer}). The following is suitable for our purposes
and postulates a hyperbolic propagation within lightlike
characteristics.

\begin{Def}\label{primitive}
The net $\{\cA(V)\}$ is said to satisfy the {\em local primitive causality} 
condition if $\cA(V'')=\cA(V)$ for every nonempty convex region $V$. 
\end{Def}

\noindent  This is an additional assumption, which does not follow from 
assumptions (i)--(vi). Indeed, there exist nets associated with certain
generalized free fields which satisfy conditions (i)--(vi) but which
violate local primitive causality \cite{Garber}. However, this condition
has been verified in many concrete models. For further insight 
into the content of local primitive causality, see the discussion directly
after the proof of Prop. 3 below.

     For later purposes we collect some definitions concerning the 
independence of local algebras.\footnote{For the origin and a 
detailed analysis of the interrelation of these and other notions of 
statistical independence, see the review \cite{Summers1990} and Chapter 
11 in \cite{konyvem} --- for more recent results, see 
\cite{Florig-Summers1997}\cite{Hamhalter}.} 
A pair $(\cA_1,\cA_2)$ of \C subalgebras of the \C algebra $\cC$
has the Schlieder property if $XY\not=0$ for any $0\not= X\in\cA_1$ and 
$0\not=Y\in\cA_2$. We note that given assumptions (i)--(v), 
$(\cA(V_1),\cA(V_2))$ has the Schlieder property for all spacelike 
separated, regular shaped\footnote{see precise 
statements in \cite{Summers1990}} $V_1,V_2$, in particular for double cones.

     A pair $(\cA_1,\cA_2)$ of such algebras is called \C independent
if for any state $\phi_1$ on $\cA_1$ and for any state $\phi_2$ on
$\cA_2$ there exists a state $\phi$ on $\cC$ such that
$\phi(X)=\phi_1(X)$ and $\phi(Y)=\phi_2(Y)$, for any
$X\in\cA_1,Y\in\cA_2$.  Under assumptions (i)--(v), algebras
associated with spacelike separated double cones are \C independent,
since, as pointed out above, they are a mutually commuting pair of
algebras satisfying the Schlieder property, which in this context is
equivalent with \C independence \cite{Roos}.

     Two von Neumann subalgebras $\cN_1,\cN_2$ of the von Neumann 
algebra $\cN$ are called logically independent 
\cite{Redei1995a}\cite{Redei1995d} if $A\es B\not=0$ for 
any  projections $0\not=A\in\cN_1$, $0\not=B\in\cN_2$. If 
$\cN_1,\cN_2$ is a mutually commuting pair of von Neumann 
algebras, then \C independence and logical independence are equivalent
\cite{konyvem}.\footnote{If $\cN_1,\cN_2$ do not mutually
commute, then \C independence is strictly weaker than logical
independence \cite{Hamhalter}.} In light of our preceding remarks, we
conclude:

\begin{lemma} \label{logind} Assumptions (i)--(v) entail that the pair
$(\cA(V_1),\cA(V_2))$ is logically independent for any spacelike 
separated double cones $V_1,V_2$.
\end{lemma}

     Let $V_1$ and $V_2$ be two spacelike separated spacetime 
regions and $A\in\cA(V_1)$ and $B\in\cA(V_2)$ be two projections. If $\phi$ is 
a state on $\cA(V_1\cup V_2)$, then it can happen that 
\begin{equation}\label{supcorr}
\phi(A\es B)>\phi(A)\phi(B) \quad .
\end{equation}
If (\ref{supcorr}) is the case, then we say that there is {\em superluminal 
(or spacelike) correlation} between $A$ and $B$ in the state $\phi$. 
We now explain why such correlations are common when assumptions
(i)--(v) hold.

     The ubiquitous presence of superluminal correlations is one of the
consequences of the generic violation of Bell's inequalities in AQFT. 
To make this clear, we need to establish some further notions.
Let $\cN_1$ and $\cN_2$ be two commuting von Neumann subalgebras of the von 
Neumann algebra $\cN$ and let $\phi$ be a state on $\cN$. 
Following \cite{Summers-Werner1985}, the Bell correlation 
$\beta(\phi,\cN_1,\cN_2)$ between the algebras $\cN_1,\cN_2$ 
in state $\phi$ is defined by
\begin{equation} \label{defbellcorr}
\beta(\phi,\cN_1,\cN_2)\equiv
\sup \;
\frac{1}{2}\phi(X_1(Y_1+Y_2)+X_2(Y_1-Y_2)) \quad ,
\end{equation}
where the supremum in (\ref{defbellcorr}) is taken over all self--adjoint 
$X_i\in\cN_1,Y_j\in\cN_2$ with norm less than or equal to 1.  It can 
be shown \cite{Cirelson}\cite{Summers-Werner1987b} that 
$\beta(\phi,\cN_1,\cN_2)\leq\sqrt{2}$. The Clauser--Holt--Shimony--Horne
version of Bell's inequality in this notation reads: 
\begin{equation}\label{defbellineq}
 \beta(\phi,\cN_1,\cN_2)\leq 1
\quad ,
\end{equation} 
and a state $\phi$ for which $\beta(\phi,\cN_1,\cN_2)> 1$ is called {\em Bell 
correlated}. It is known \cite{Summers-Werner1987b} that if $\cN_1$ or 
$\cN_2$ is abelian, or if $\phi$ is a product state across the algebras 
$\cN_1,\cN_2$ ({\it i.e.}, if $\phi(XY) = \phi(X)\phi(Y)$, for all 
$X \in \cN_1$ and $Y \in \cN_2$), then $\beta(\phi,\cN_1,\cN_2)= 1$. 
Hence, if $\phi$ is Bell correlated then $\phi$ cannot be a product
state across the algebras $\cN_1,\cN_2$. But, as we shall next see, 
non--product states yield correlations (\ref{supcorr}) for suitable
choices of projections.

\begin{lemma}\label{nonproduct} 
Let $\cN_1$ and $\cN_1$ be subalgebras of the von
Neumann algebra $\cN$ such that $\cN_1 \subset \cN_2 '$, and let
$\phi$ be a normal state on $\cN$ which is not a product state across the
algebras $\cN_1,\cN_2$. Then there exist projections $A \in \cN_1$
and $B \in \cN_2$ such that $\phi(A\es B) > \phi(A)\phi(B)$.
\end{lemma}

\begin{proof} By the spectral theorem, if 
$\phi(A\es B)=\phi(A)\phi(B)$, for all projections $A\in\cN_1$ and 
$B\in\cN_2$, then $\phi$ is a product state across $\cN_1,\cN_2$. 
Hence, there must exist projections $A \in \cN_1$ and $B \in \cN_2$ such
that $\phi(A\es B)\neq\phi(A)\phi(B)$. If $\phi(A\es
B)<\phi(A)\phi(B)$, then
\begin{eqnarray*} 
\phi((1-A)B) & = & \phi(B) - \phi(AB) \, > \, \phi(B) - \phi(A)\phi(B) \\
    & = & \phi(1-A)\phi(B) \quad ,
\end{eqnarray*}
and $1-A \in \cN_1$ is a projector. Similarly, 
$\phi(A(1-B))>\phi(A)\phi(1-B)$. On the other hand, if
$\phi(A\es B)>\phi(A)\phi(B)$, then also
$\phi((1-A)\es (1-B))>\phi(1-A)\phi(1-B)$. 
\end{proof}

\medskip

     There are many situations in which 
$\beta(\phi,\cA(V_1),\cA(V_2)) = \sqrt{2}$, in other words, where
there is maximal violation of Bell's inequalities 
(cf. \cite{Summers-Werner1987b}\cite{Summers-Werner1987a}\cite{Summers-Werner1988}). Indeed,
if $V_1$ and $V_2$ are tangent spacelike separated wedges or double
cones, there is maximal violation of Bell's inequalities in {\em  every} 
normal state \cite{Summers-Werner1988}. We refer the interested reader 
to those papers. To keep the discussion as simple as
possible, we shall only discuss the recent results established by Halvorson
and Clifton.\footnote{In point of fact, Halvorson and Clifton required of 
$\cN_1$ and $\cN_2$ that they be of infinite type, not the more restrictive
type $III$, but this is the
version which we shall employ below.} The symbol $\cN_1\vee\cN_2$ denotes
the smallest von Neumann algebra containing both $\cN_1$ and $\cN_2$.

\begin{Proposition}\label{manycorr}  {\rm \cite{Halvorson-Clifton}}
If $(\cN_1,\cN_2)$ is a pair of commuting type $III$ von Neumann 
algebras acting on the Hilbert space $\cH$ and having the Schlieder property, 
then the set of unit vectors which induce Bell correlated states on 
$\cN_1,\cN_2$ is open and dense in the unit sphere of $\cH$. Indeed,
the set of normal states on $\cN_1\vee\cN_2$ which are Bell correlated 
on $\cN_1,\cN_2$ is norm dense in the normal state space of $\cN_1\vee\cN_2$.
\end{Proposition}
From the above remarks, given the assumptions (i)--(vi), we see that
for any spacelike separated double cones $V_1,V_2$, the pair
$(\cA(V_1),\cA(V_2))$ satisfies the hypothesis of
Prop. \ref{manycorr}.  So, ``most'' normal states on such pairs of algebras
manifest superluminal correlations (\ref{supcorr}).\footnote{If 
$V_1$ and $V_2$ are tangent spacelike separated wedges or double cones, 
{\it all} normal states on $(\cA(V_1),\cA(V_2))$ manifest superluminal 
correlations.} Hence, superluminal correlations abound in AQFT, and
the question posed in the introduction is not vacuous. In the next
section, we address this question.

\section{The Notion of Reichenbachian Common Cause in AQFT}

Let $(\Omega,p)$ be a classical probability measure space with Boolean algebra 
$\Omega$ and probability measure $p$.
If $A,B\in\Omega$ are such that 
\begin{equation} \label{corr}
p(A\es B)>p(A)p(B) \quad ,
\end{equation}
then the events $A$ and $B$ are said to be (positively) {\em correlated}. 

\begin{Def} \label{cc}
$C\in\Omega$ is a 
common cause of the correlation (\ref{corr}) if the following (independent) 
conditions hold: 

\begin{eqnarray}
p(A\es B|C)&=&p(A|C)p(B|C) \quad ,  \label{screenoff1}\\
p(A\es B|C^\c)&=&p(A|C^\c)p(B|C^\c) \quad ,  \label{screenoff2}\\
p(A|C)&>&p(A|C^\c) \quad ,  \label{nagyobb1}\\
p(B|C)&>&p(B|C^\c) \quad , \label{nagyobb2}
\end{eqnarray}
where $p(X|Y)$ denotes here the conditional probability of $X$ 
on condition $Y$, and it is assumed that none of the probabilities $p(X)$, 
$(X=A,B,C)$ is equal to zero. 
\end{Def}

     The above definition is due to Reichenbach
(\cite{Reichenbach1956}, Section 19). We wish to extend this
definition to the setting of AQFT. To do this, we first define a
notion of common cause of a correlation in a noncommutative measure
space $(\PN,\phi)$ with a non-distributive von Neumann lattice $\PN$
in place of the Boolean algebra $\Omega$ and a normal state $\phi$
playing the role of $p$.\footnote{Only normal states are considered,
since they have the continuity property which generalizes the property
of $\sigma$--additivity in the classical context \cite{KadRing}.}
Here, $\cN$ is a von Neumann algebra and $\PN$ denotes the set of
projections in $\cN$.

\begin{Def}\label{defqcc}
Let $A,B\in\PN$ be two commuting projections which are correlated in $\phi$:
\begin{equation}\label{qcorr}
\phi(A\es B)>\phi(A)\phi(B) \quad .
\end{equation} 
$C\in\PN$ is a {\em common cause} of the correlation (\ref{qcorr}) if
$C$ commutes with both $A$ and $B$ and the following conditions (completely 
analogous to (\ref{screenoff1})-(\ref{nagyobb2})) hold:
\begin{eqnarray} {\phi(A\es B\es 
C)\over\phi(C)}&=&{\phi(A\es C)\over\phi(C)}{\phi(B\es C)\over\phi(C)} \quad ,
\label{AQFTscreen1}\\ 
{\phi(A\es B\es C^\c)\over\phi(C^\c)}&=&{\phi(A\es 
C^\c)\over\phi(C^\c)} {\phi(B\es C^\c) \over\phi(C^\c)} \quad ,
\label{AQFTscreen2}\\ 
{\phi(A\es C)\over\phi(C)}&>&{\phi(A\es C^\c)\over\phi(C^\c)} \quad , 
\label{AQFTnagy1}\\ 
{\phi(B\es C)\over\phi(C)}&>&{\phi(B\es C^\c)\over\phi(C^\c)} \quad .
\label{AQFTnagy2} 
\end{eqnarray} 
\end{Def}
 
     Definition \ref{defqcc} is a natural specification in a noncommutative 
probability space $(\PN,\phi)$ of the classical notion of common cause. The 
only deviation from Definition \ref{cc} is that the commutativity of the 
events involved is now  required explicitly. One could, in principle, also 
define a common cause $C$ which does not commute with $A$ or $B$, but then one 
would have to expand Reichenbach's original scheme by allowing noncommutative 
conditionalization, which we do not wish to consider here. (See the papers
\cite{Szabogabor1}, \cite{Szabogabor2} for an analysis of some technical and 
conceptual difficulties concerning the generalization of Reichenbach's scheme 
to non-distributive event structures.) 

     The following formulation of Reichenbach's Common Cause Principle 
for local relativistic nets was proposed in \cite{Redei1997}.

\begin{Def} \label{defccqft} Let $\{\cA(V)\}$ be a net of local von 
Neumann algebras over Minkowski space. Let $V_1$ and $V_2$ be two 
spacelike separated 
spacetime regions, $BLC(V_1)$ and $BLC(V_2)$ be their backward light 
cones, and let $\phi$ be a locally normal state on the quasilocal 
algebra $\cA$.\footnote{A state $\phi$ on $\cA$ is locally normal if 
its restriction to $\cA(V)$ is normal for every double cone $V$.} 
We say that the local system $(\cA(V_1),\cA(V_2),\phi)$ 
{\em satisfies Reichenbach's Common Cause Principle} if and only if
for any pair of projections $A\in\cA(V_1)$ $B\in\cA(V_2)$ we have the 
following: if 
\begin{equation}\label{qftcorr} 
\phi(A\es B)>\phi(A)\phi(B) \quad ,
\end{equation} 
then there exists a projection $C$ in the von Neumann algebra $\cA(V)$  
associated with a region $V$ such that 
$$ 
V\subseteq (BLC(V_1)\setminus V_1)\cap (BLC(V_2)\setminus V_2) \quad 
$$ 
and such that $C$ is a common cause of the correlation (\ref{qftcorr})
in the sense of Definition \ref{defqcc}.
We say that Reichenbach's Common Cause Principle holds for the net 
iff for every pair of spacelike separated spacetime regions $V_1,V_2$ and 
every normal state $\phi$, Reichenbach's Common Cause Principle holds 
for the local system $(\cA(V_1),\cA(V_2),\phi)$. 
\end{Def}

     As we indicated in the Introduction, it is still an open problem
whether this condition holds for the nets of local von Neumann
algebras occurring in AQFT. The next definition formulates a weaker
form of the Common Cause Principle for a net $\{\cA(V)\}$ --- it is
for this weaker notion that we can prove something in this paper.

\begin{Def}\label{wdefccqft}
We say that the local system $(\cA(V_1),\cA(V_2),\phi)$ satisfies 
the {\em Weak Reichenbach's Common Cause Principle} 
if and only if for any pair of 
projections $A\in\cA(V_1)$ $B\in\cA(V_2)$ we have the following: if 
\begin{equation} \label{qcorr2}
\phi(A\es B)>\phi(A)\phi(B) \quad ,
\end{equation} 
then there exists a projection $C$ in the von Neumann algebra $\cA(V)$  
associated with a region $V$ such that 
\begin{displaymath} 
V\subseteq (BLC(V_1)\setminus V_1)\cup (BLC(V_2)\setminus V_2) 
\end{displaymath} 
and such that $C$ is a common cause of the correlation (\ref{qcorr2}) in the 
sense of Definition \ref{defqcc}.
\end{Def}

\begin{Proposition}\label{main}
If the net $\{\cA(V)\}$ satisfies conditions (i)--(vi) and local 
primitive causality, then every local system $(\cA(V_1),\cA(V_2),\phi)$ 
with $V_1,V_2$ contained in a pair of spacelike separated double cones
and with a locally normal and locally faithful state $\phi$ 
satisfies Weak Reichenbach's Common Cause Principle.
\end{Proposition} 

\noindent
Note that because logical independence is hereditary, Lemma 1 entails that
the pair of algebras $\cA(V_1),\cA(V_2)$ in Prop. \ref{main} is logically
independent. The proof of Prop. \ref{main} is based on the following 
two Lemmas:

\begin{lemma} \label{suffcond}
Let $\phi$ be a faithful state on a von Neumann algebra $\cN$
containing two mutually commuting subalgebras $\cN_1,\cN_2$ which are
logically independent. Let $A \in \cN_1$ and $B \in \cN_2$ be projections 
satisfying (\ref{qcorr2}). Then a sufficient condition for $C$ to satisfy 
(\ref{AQFTscreen1})-(\ref{AQFTnagy2}) is that the following two conditions 
hold:
\begin{eqnarray}
C & < & A\es B \quad , \label{eleg1}  \\
\phi(C)&=&\frac{\phi(A\es B)-\phi(A)\phi(B)}{1-\phi(A\vagy B)} \quad .
\label{eleg2}
\end{eqnarray}
\end{lemma}

\begin{lemma} \label{proj}
Let $\cN$ be a type $III$ von Neumann algebra on a separable
Hilbert space $\cH$, and let $\phi$ be 
a faithful normal state on $\cN$. Then for every projection $A\in\PN$ and 
every positive real number $0<r<\phi(A)$ there exists a projection $P\in\PN$ 
such that $P<A$ and $\phi(P)=r$. 
\end{lemma}

\medskip 
\noindent 
{\bf Proof of Lemma \ref{suffcond}:}
Note first that if $A$ and $B$ are correlated, then one must have 
$1-\phi(A\vagy B)>0$. If, on the contrary, one had 
$\phi(1) = 1 = \phi(A\vagy B)$, then since 
$1 = A \vee B + A^{\bot}\wedge B^{\bot}$, one would have 
\begin{equation}\label{contr}
\phi(A^{\bot}\wedge B^{\bot}) = 0 \quad .
\end{equation}
On the other hand, since $A$ and $B$ are correlated, it follows that
$$
\phi(A^{\bot}\wedge B^{\bot}) > \phi(A^{\bot})\phi(B^{\bot}) \quad ,
$$
contradicting (\ref{contr}).
Hence, the right hand side of (\ref{eleg2}) is well defined 
and greater than zero. Further note that since $A \geq A \es B$ and
$B \geq A \es B$, one must have $\phi(A) \geq \phi(A \es B)$ and
$\phi(B) \geq \phi(A \es B)$. But equality in both of these cases is
excluded by hypothesis, since if $\phi(A)=\phi(A\es B)$ or 
$\phi(B)=\phi(A\es B)$, then by the faithfulness of $\phi$ one has 
$A=A\es B$ or $B=A\es B$. This would imply $B\geq A$ or $A\geq B$, and so 
$A\es B^{\bot}=0$ or $B\es A^{\bot}=0$, which
contradicts the logical independence of the pair $\cN_1,\cN_2$.
Hence, one has $\phi(A) - \phi(A \es B) \equiv a > 0$ and
$\phi(B) - \phi(A \es B) \equiv b > 0$.

     Elementary algebraic calculation shows that the inequality
\begin{displaymath}
\frac{\phi(A\es B)-\phi(A)\phi(B)}{1-\phi(A\vagy B)}<\phi(A\es B)
\end{displaymath}
is equivalent to the inequality
\begin{equation} \label{detail}
\phi(A)\phi(B) > [\phi(A) + \phi(B) - \phi(A \es B)]\,\phi(A \es B) \quad .
\end{equation}
But 
\begin{displaymath}
\phi(A)\phi(B) = \phi(A)[b + \phi(A \es B)]
\end{displaymath}
and
\begin{displaymath}
[\phi(A) + \phi(B) - \phi(A \es B)]\phi(A \es B) = [\phi(A) + b]\,\phi(A \es B)
\quad .
\end{displaymath}
Since $\phi(A)b > b\phi(A \es B)$, inequality (\ref{detail}) follows.
Therefore, the right hand side of (\ref{eleg2}) --- 
and hence the value of $\phi(C)$ --- is smaller than $\phi(A\es B)$, 
and so the conditions (\ref{eleg1}) and (\ref{eleg2}) are compatible. 

     It remains to see that conditions (\ref{AQFTscreen1})-(\ref{AQFTnagy2}) 
hold. Conditions (\ref{AQFTscreen1}), (\ref{AQFTnagy1}) and (\ref{AQFTnagy2}) 
hold trivially, and since $C< A\es B$ commutes with both $A$ and $B$, one can 
write 
\begin{displaymath}
\phi(X)=\phi(X\es C)+\phi(X\es C^{\bot}) \quad , \quad X=A,B,A \es B \quad ,
\end{displaymath}
implying 
\begin{displaymath}
\phi(X\es C^{\bot})=\phi(X)-\phi(X\es C)=\phi(X)-\phi(C) \quad , \quad
X=A,B, A \es B \quad .
\end{displaymath}
Hence, (\ref{AQFTscreen2}) is equivalent to
\begin{displaymath}
\frac{\phi(A\es B)-\phi(C)}{\phi(C^{\bot})}=
\frac{\phi(A)-\phi(C)}{\phi(C^{\bot})}
\frac{\phi(B)-\phi(C)}{\phi(C^{\bot})} \quad , 
\end{displaymath}
which, in turn, is equivalent to
\begin{displaymath}
\phi(C^{\bot})\lbrack\phi(A\es B)-\phi(C)\rbrack =
\lbrack\phi(A)-\phi(C)\rbrack\lbrack\phi(B)-\phi(C)\rbrack \quad .
\end{displaymath}
This latter equation can be seen after elementary algebra to be 
equivalent to (\ref{eleg2}).
{\hfill $\square$}

\bigskip

\noindent
{\bf Proof of Lemma \ref{proj}:} Consider the set $S$ of projections defined by
\begin{displaymath}
S\equiv\{X\in\PN \ \vert\  X<A \ \mbox{and\ }\phi(X)\leq r\} \quad .
\end{displaymath}
$S$ is partially ordered with respect to the partial 
ordering inherited from the standard lattice ordering in $\PN$. 
Let $S'$ be a linearly ordered subset of $S$. Since $\PN$ is a 
complete lattice, the least upper bound $Z$ of $S'$ exists. Since $Z$ is 
the least upper bound of $S'$, one has $Z\leq A$, and since $Z$ is the 
strong operator limit of the net $S'$ and $\phi$ is normal, one also has 
$\phi(Z)\leq r$. This implies that $Z<A$ holds, as well 
(because $r<\phi(A)$ and $\phi$ is faithful). So every linearly 
ordered subset of $S$ has a maximal element in $S$; therefore by 
Zorn's Lemma, $S$ has a maximal element $P$. 

     It shall be shown that $\phi(P)=r$. Assume for the purpose of
contradiction that $\phi(P)<r$ and consider $Q\equiv A\es P^{\bot}\in\PN$. 
Since $P<A$, it follows that $Q\not=0$. Since $\cN$ is a type $III$ algebra
and $\cH$ is separable, there exist a countably infinite number of 
mutually orthogonal projections $Q_i\in\PN$, each equivalent to $Q$ in 
the Murray-von Neumann sense (cf. \cite{KadRing}), such that 
$Q=\vagy_i Q_i$. So one has\footnote{The second equality in the display 
is precisely the generalization of $\sigma$--additivity to which we 
have previously alluded and which obtains for normal states, but not 
for general states.} 
\begin{displaymath}
\phi(Q) = \phi(\vagy_i Q_i)=\sum_i\phi(Q_i) \quad .
\end{displaymath}
Since $\phi$ is faithful, one must have $\phi(Q_i)>0$ for all $i$, and so 
$\phi(Q_i)\to 0$ ($i\to\infty$). Consequently, there exists a sufficiently
large $k$ such that for the projection $P'\equiv P+Q_k$ one has 
\begin{displaymath}
r \, > \, \phi(P') = \phi(P+Q_k)=\phi(P)+\phi(Q_k)>\phi(P) \quad ,
\end{displaymath}
and $P < P' < A$, which contradicts the maximality of $P$
in $S$. 
{ \hfill $\square$}

\bigskip

\noindent
{\bf Proof of Proposition \ref{main}:}  
Let $\phi$ be a locally normal and locally faithful state on $\cA$ 
and assume that the projections $A\in\cA(V_1),B\in\cA(V_2)$ are 
correlated in $\phi$: 
\begin{displaymath}
\phi(A\es B)>\phi(A)\phi(B) \quad .
\end{displaymath} 
Since $V_1$ and $V_2$ are bounded, there exists a bounded (hyper)rectangular
spacetime region $V$ such that
\begin{displaymath}
V\subseteq (BLC(V_1)\setminus V_1)\cup (BLC(V_2)\setminus V_2)
\end{displaymath}
and such that the causal completion $V''$ (which is a double cone) 
of $V$ contains $V_1\cup V_2$. Hence,  by 
isotony and local primitive causality one has 
\begin{displaymath}
A,B \in \cA(V_1\cup V_2)\subseteq \cA(V'')=\cA(V) \quad .
\end{displaymath}
Let 
\begin{displaymath}
r\equiv \frac{\phi(A\es B)-\phi(A)\phi(B)}{1-\phi(A\vagy B)}>0 \quad .
\end{displaymath}
It has already been established in the proof of Lemma \ref{suffcond} that 
$r < \phi(A \es B)$. Since double cone algebras are type $III$, by 
Lemma \ref{proj} there exists a projection $C\in\cA(V)$ such that 
$C<A\es B$ and $\phi(C)=r$, and by Lemma \ref{suffcond} this $C$ 
satisfies conditions (\ref{AQFTscreen1})-(\ref{AQFTnagy2}). 
{ \hfill $\square$ }

\bigskip

     The observant reader may be disturbed to note that in the proof
just given the projections $A,B$ were initially localized in the
algebra $\cA(V_1 \cup V_2)$ and then were located in the algebra
$\cA(V)$, even though, in fact, we have seen that 
$V \cap (V_1 \cup V_2) = \emptyset$.  
What kind of sleight of hand has taken place here?
It is not widely understood that an ``observable'' $A$ does not
represent a unique measuring apparatus in some fixed laboratory, but
rather represents an equivalence class of such apparata
(cf. \cite{Neumann-Werner}).  Consider two such idealized apparata
$X,Y$ such that $\phi(X) = \phi(Y)$ for all (idealized) states
$\phi$ admitted in the theory (the set of such states contains as
a subset --- at least in principle --- all states preparable in the
laboratory). These two apparata are then identified to be in the same
equivalence class and are thus represented by a single operator $A \in \cA$. 
Hence, each of the projections $A,B$ above, which are localized
simultaneously in $V$ and $V_1 \cup V_2$, represents two distinct
events --- one taking place in $V$ and the other taking place in 
$V_1 \cup V_2$. The fact that it is possible, for every event in 
$V_1 \cup V_2$, to find an event in $V$ which is equivalent to the first
in the stated sense is part of the content of the local primitive causality
condition.  With this in mind, the reader is in the position to better
appreciate the significance of the fact that the primitive causality
condition can, in fact, be deduced for nets locally associated with,
for example, free quantum fields satisfying hyperbolic equations of
motion, as well as some interacting quantum field models whose
construction has been carried out with mathematical rigor.

     To better understand the nature of the causal requirement being
made here, we recall another, less frequently used algebraic causality
condition. We say that $V_1$ is in the causal shadow of $V_2$ if every
backward light ray from any point of $V_1$ passes through $V_2$. Then
a more explicit algebraic formulation of the physical requirement that
there is some kind of hyperbolic propagation of physical effects in
action is that $\cA(V_1) \subset \cA(V_2)$, whenever $V_1$ is in the
causal shadow of $V_2$. It is easy to see that the local primitive
causality condition implies the latter condition, and it is clear that 
this latter causality condition would suffice to entail the conclusion of
Prop. 3.
 
\section{Concluding Remarks}

Proposition \ref{main} does not give an answer to the question of whether 
AQFT satisfies Reichenbach's Common Cause Principle interpreted in the sense 
of Definition \ref{defccqft}, because it locates the common cause $C$ 
only within 
the union of the backward light cones of $V_1$ and $V_2$ rather than in the 
intersection of these light cones. A bit more, however, can 
be said of the location of the specific common cause $C$ displayed in the 
proof of 
Proposition \ref{main}. 
Define $\tilde{V}_1$ and $\tilde{V}_2$ by
\begin{eqnarray}
\tilde{V}_1&\equiv& (BLC(V_1)\cap V)\setminus (BLC(V_1)\cap BLC(V_2)) \\
\tilde{V}_2&\equiv& (BLC(V_2)\cap V)\setminus (BLC(V_1)\cap BLC(V_2)) 
\end{eqnarray}
(i.e. $\tilde{V}_1$ and $\tilde{V}_2$ are the parts of $V$ that are in the 
backward light cones of $V_1$ and $V_2$, respectively, 
but do not intersect with the part of 
$V$ which is in the common causal past of $V_1$ and $V_2$). 
Since $(\tilde{V}_1\cup V_1)$ and $(\tilde{V}_2\cup V_2)$ are contained in
spacelike separated double cones, the algebras $\cN(\tilde{V}_1\cup V_1)$ and 
$\cN(\tilde{V}_2\cup V_2)$ are logically independent, hence the common cause 
$C< A\es B$ cannot belong to $\cN(\tilde{V}_1)$ or to $\cN(\tilde{V}_2)$ 
only, so neither $V\subseteq \tilde{V}_1$ nor $V\subseteq \tilde{V}_2$ 
is possible.

     The common cause $C< A\es B$ is a very specific one --- it {\it implies} 
both $A$ and $B$. Such a common cause was called in \cite{Redei1997} a 
{\em strong} common cause, whereas a common cause $C$ is called 
{\em genuinely probabilistic} if neither $C\leq A$ nor $C\leq B$ is the 
case. We conjecture that the Weak Reichenbach's Common Cause Principle 
holds also with a genuinely probabilistic common cause. Indeed, we expect
there to be an extraordinary richness of common causes of this type
for any pair of correlated projections. After all, this is already
evident for the strong common causes whose existence we have already
established. A little thought will make clear that there is an infinite
number of mutually disjoint (hyper)rectangular regions $V \subset M$
which are contained in $BLC(V_1) \cup BLC(V_2)$, disjoint from $V_1 \cup V_2$ 
and which satisfy $V_1 \cup V_2 \subset V''$. Hence, for given correlated
projections $A,B$ as described, there are infinitely many different
strong common causes.

     In \cite{bjps} the probability space $(\Omega,p)$ was called 
{\em common cause incomplete} if it contains a pair of events $(A,B)$ 
which are correlated in $p$ but for which $\Omega$ does not contain a common 
cause $C$ of the correlation between $A$ and $B$. It was shown in 
\cite{bjps} that every common cause incomplete probability space can 
be extended in such a way that the extension contains a common cause 
of the given correlated pair. Common cause incompleteness of a 
noncommutative space $(\PN,\phi)$ can be defined in complete analogy with the 
classical case (taking Definition \ref{defqcc} as the definition of common 
cause), and it can be shown \cite{bjps}, \cite{kozosijtp} that common cause 
incomplete noncommutative spaces also can be extended in such a manner that 
the extension contains a common cause of a given correlation. Because of 
conditions (\ref{nagyobb1})--(\ref{nagyobb2}) 
(respectively (\ref{AQFTnagy1})--(\ref{AQFTnagy2})) required of the common 
cause, extending a given $(\Omega,p)$ (respectively $(\PN,\phi)$) by 
``adding'' a common cause to it entails that the extension contains 
correlations which are not present in the original structure. It is therefore 
a nontrivial matter whether a given common cause incomplete space, classical 
or quantum, can be made {\em common cause closed}, or whether common cause 
closed spaces exist at all, where ``common cause closedness'' of a probability 
space means that the space contains a common cause of {\em every} correlation 
in it. It can be shown \cite{Gyenis-Redei} that while common cause closed 
classical probability spaces exist, they cannot be small --- no 
$(\Omega,p)$ with a finite Boolean algebra $\Omega$ can be common cause 
closed, with the exception of the Boolean algebra $\Omega$ generated 
by 5 atoms. 

     But the requirement of common cause closedness is too 
restrictive on intuitive grounds as well. One does not expect to have a 
common cause explanation of probabilistic correlations which arise 
as a consequence of a direct physical influence between the correlated 
events, or which are due to some logical relations between the correlated 
events. Thus it is a more reasonable to demand a space $(\PN,\phi)$ to be 
common cause closed with respect to two logically independent von Neumann 
sub-lattices ${\cal P}(\cN_1)$ and ${\cal P}(\cN_1)$ in $\PN$. Clearly, 
Lemma \ref{suffcond} and Lemma \ref{proj} imply that $(\PN,\phi)$ is 
common cause closed with respect to any two logically independent von 
Neumann sublattices, if $\cN$ is a type $III$ algebra and $\phi$ is a 
faithful normal state on $\cN$. 

\bigskip
\noindent
{\bf Acknowledgement} Work supported by OTKA (contract numbers T
032771 and T 035234).

\end{document}